%% file: xe100_screening_paper.tex
\documentclass[final,5p,times,twocolumn]{elsarticle}

\usepackage{epsfig}
\usepackage{graphicx}
\usepackage{rotating}
\usepackage{supertabular}
\usepackage{longtable}
\usepackage{lscape}
\usepackage{multirow}
\usepackage{multicol}
\usepackage[left]{lineno}
\usepackage{color}

\usepackage{bm}
\usepackage{dblfloatfix}

\long\def\symbolfootnote[#1]#2{\begingroup\def\thefootnote{\fnsymbol{footnote}}\footnote[#1]{#2}\endgroup}

\newenvironment{zenumerate}{\newcounter{zitem}\setcounter{zitem}{0}}
{}
\newcommand\zitem{\refstepcounter{zitem}\thezitem. }

\newenvironment{aenumerate}{\newcounter{aitem}\setcounter{aitem}{0}}
{}
\newcommand\aitem{\refstepcounter{aitem}\theaitem. }

\journal{???}

\begin{document}

\begin{frontmatter}

\title{Material screening and selection for XENON100}

\author[add:columbia]{E.~Aprile}
\author[add:ucla]{K.~Arisaka}
\author[add:lngs]{F.~Arneodo}
\author[add:zurich]{A.~Askin}
\author[add:zurich]{L.~Baudis}
\author[add:zurich]{A.~Behrens}
\author[add:muenster]{K.~ Bokeloh}
\author[add:ucla]{E.~Brown}
\author[add:coimbra]{J.M.R.~Cardoso}
\author[add:columbia]{B.~Choi}
\author[add:ucla]{D.~Cline}
\author[add:lngs,add:mainz]{S.~Fattori}
\author[add:zurich]{A.D.~Ferella}
\cortext[cor1]{ferella@physik.uzh.ch}
\author[add:columbia]{K.L.~Giboni}
\author[add:zurich]{A.~Kish}
\author[add:ucla]{C.W.~Lam}
\author[add:subatech]{J.~Lamblin}
\author[add:columbia]{R.F.~Lang}
\author[add:columbia]{K.E.~Lim}
\author[add:coimbra]{J.A.M.~Lopes}
\author[add:zurich]{T.~Marrod\'an Undagoitia}
\author[add:rice]{Y.~Mei}
\author[add:columbia]{A.J.~Melgarejo Fernandez}
\author[add:shanghai]{K.~Ni}
\author[add:mainz,add:rice]{U.~Oberlack}
\author[add:coimbra]{S.E.A.~Orrigo}
\author[add:ucla]{E.~Pantic}
\author[add:columbia]{G.~Plante}
\author[add:coimbra]{A.C.C~Ribeiro}
\author[add:zurich]{R.~Santorelli}
\author[add:coimbra]{J.M.F.~dos~Santos}
\author[add:zurich,add:rice]{M.~Schumann}
\author[add:rice]{P.~Shagin}
\author[add:ucla]{A.~Teymourian}
\author[add:subatech]{D.~Thers}
\author[add:zurich]{E.~Tziaferi}
\author[add:ucla]{H.~Wang}
\author[add:muenster]{C.~Weinheimer}
\author{\\(XENON100 Collaboration)}
\author[add:lngs]{\\M.~Laubenstein}
\author[add:lngs]{S.~Nisi}

\address[add:columbia]{Department of Physics, Columbia University, New
York, NY 10027, USA}
\address[add:zurich]{Physik-Institut, Universit\"at Z\"urich, 8057
Z\"urich, Switzerland}
\address[add:lngs]{INFN -- Laboratori Nazionali del Gran Sasso, 67010
Assergi, Italy}
\address[add:coimbra]{Department of Physics, University of Coimbra,
R.~Larga, 3004-516, Coimbra, Portugal}
\address[add:rice]{Department of Physics \& Astromony, Rice University,
Houston, TX, 77251, USA}
\address[add:ucla]{Department of Physics \& Astromony, University of
California, Los Angeles, CA, 90095, USA}
\address[add:shanghai]{Shanghai Jiao Tong University, Shanghai, China}
\address[add:muenster]{Institut f\"ur Kernphysik, Universit\"at M\"unster,
48149 M\"unster, Germany}
\address[add:mainz]{Institut f\"ur Physik, Johannes Gutenberg-Universit\"at Mainz, 55099 Mainz, Germany}
\address[add:subatech]{SUBATECH, Ecole des Mines de Nantes, Universit\'e
de Nantes, CNRS/IN2P3 Nantes, France}

\begin{abstract}
Results of the extensive radioactivity screening campaign to identify
materials for the construction of XENON100 are reported. This Dark
Matter search experiment is operated underground at Laboratori Nazionali
del Gran Sasso (LNGS), Italy. Several ultra sensitive High Purity Germanium
detectors (HPGe) have been used for gamma ray spectrometry. Mass spectrometry 
has been applied for a few low mass plastic samples. Detailed tables with the radioactive contaminations
of all screened samples are presented, together with the implications for XENON100.
\end{abstract}


\end{frontmatter}

\section{Introduction}

The indirect evidence of a significant cold dark matter component in the Universe
\cite{freedman2003,bullet_cluster,dm_ring} motivated the start
of several experiments aiming to directly detect Dark Matter 
in the form of Weakly Interacting Massive Particles (WIMPs)
\cite{gaitskell2004,gabriel05,laura06}. The possible existence of WIMPs
is supported by beyond-Standard Model theories such as supersymmetric theories (SUSY), models with
extra dimensions and little Higgs models
\cite{Bottino,Ellis,Cheng,Birkedal-Hansen}.
Several experiments are aiming to directly detect WIMP dark matter by searching for the 
elastic scattering of WIMPs
off target nuclei. XENON100, which has recently published one of the best limits
on spin-independent WIMP-nucleon scattering cross sections \cite{xe100_prl_first,xe100::pl},  
is one of the most sensitive experiments of the current generation. 
It is a double-phase (liquid/gas) time projection
chamber (TPC) \cite{LXeDouble}, which allows fiducialization of the active target and 
discrimination of the nuclear
recoil signal from electronic recoils, induced by background radiation. The former is made
possible by a reconstruction of the interaction vertex in three dimensions, whereas the latter is possible
because of the different charge to scintillation light ratio for the two types of 
interaction \cite{ref::aprile2006}.

In order to reach its design sensitivity, the experiment is to be built from materials with
very low intrinsic radioactivity. Therefore, the radioactive screening and selection of
materials has played an important role in the design and during the construction
phase of XENON100, as well as in the background simulation of the experiment \cite{Xe100BG}.
The XENON collaboration used the High Purity Germanium (HPGe) detector Gator
\cite{gator_paper},  operated by the University of Zurich, to perform gamma ray
spectrometry in order to determine the intrinsic radioactivity of materials considered
for the construction of XENON100. Moreover, HPGe detectors of the low-level
counting facility at Laboratori Nazionali del Gran Sasso (LNGS), Italy,
\cite{Arpesella} were also used. This facility includes the GeMPI-I and  GeMPI-II
detectors, which are the most sensitive low-radioactivity HPGe detectors in the
world \cite{GeMPI}. The radioactivity of a few low-mass samples has been
determined by mass spectroscopy.

This paper is organized as follows: First the radiopurity requirements of
experiments searching for Dark Matter are reviewed, followed by a short description of the
methods used for the sample screening in Sections~\ref{sec::gammaray} and~\ref{icpsm}.
The results are presented and discussed in~Section ~\ref{sec::results}.

\section{Radiopurity Requirements for Direct Dark Matter Searches and XENON100}\label{sec:radio}

Experiments searching for Dark Matter are often limited by background radiation 
from the detector materials and surroundings, including a possible radiation shield,
and by interactions induced by cosmic rays, especially muons.

Generally, particles can interact either with the atomic electrons of the detector material
(electronic recoils) or with the target nuclei (nuclear recoils). Since the electrically
neutral WIMPs and neutrons are both expected to produce nuclear recoils, neutrons are the most dangerous
background for Dark Matter experiments, as they can potentially mimic a WIMP signal.
In most cases, however, the dominating background is from electronic recoils.
For this reason, most experiments employ at least one technique to discriminate
between nuclear and electronic recoils.

The dominant electronic recoil background usually comes from $\gamma$-rays from radioactive isotopes
in the shield and in the detector itself. $\beta$-decays only
contribute when they occur in the target or at the target's surface. Depending on
the energy of the electrons, they can generate Bremsstrahlung in the vicinity of the sensitive volume,
which might add an additional source of background. $\alpha$-decays
do not directly contribute to the background at low energies,
given the typical $\alpha$ energies of $>3$~MeV. However, they might indirectly contribute to the
nuclear recoil background, as they can produce neutrons in $(\alpha,n)$ reactions and 
cause nuclear recoils from daughter nuclei.  Another source of neutrons is 
from spontaneous fission of $^{238}$U. 

The most relevant sources of radioactive contamination in the XENON100 experiment are
primordial radionuclides ($^{238}$U, $^{232}$Th and $^{40}$K),
anthropogenic radionuclides ($^{137}$Cs, $^{85}$Kr), cosmogenic radionuclides (mainly $^{60}$Co)
and environmental radioactive noble gases, such as $^{222}$Rn and $^{220}$Rn which are
daughters of the $^{238}$U and $^{232}$Th decay chains. 

\begin{figure*}[h]
\begin{center}
\includegraphics[width=0.78\textwidth]{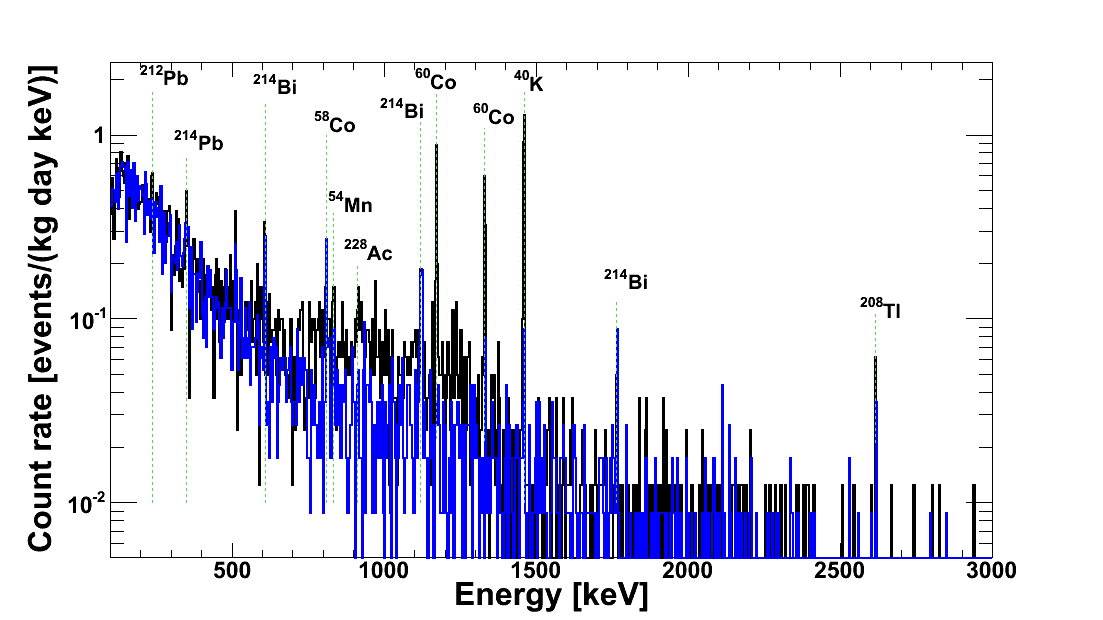}
\caption{Example of a measurement with Gator: 7~{\it Hamamatsu} R8520 PMTs were screened (entry~34 in Table~\ref{tab:screen}). The lines from their intrinsic radioactive contaminations (solid black) are clearly visible  above the Gator background spectrum (blue). }
\label{fig::r8520}
\end{center}
\end{figure*}

The goal of the XENON100 experiment is to probe spin-independent WIMP-nucleon
scattering cross sections down to the level of $\sim 2 \times 10^{-45}$~cm$^2$.
In order to reach this sensitivity, one of the key requirements on the experiment	 design
is the suppression of background and
a total electronic recoil background of $< 10^{-2}$~events~keV$^{-1}$~kg$^{-1}$~day$^{-1}$
is required. This is achieved by shielding the detector from environmental
radioactivity with a passive shield, by the detector design using an active LXe veto scintillator, fiducial volume cuts, 
and by constructing the detector from materials with a low intrinsic radioactivity \cite{Xe100BG}.

\section{Gamma Ray Spectrometry with HPGe Detectors}\label{sec::gammaray}

When measuring the intrinsic radioactivity of a sample material 
with high purity germanium (HPGe) detectors, one exploits the 
excellent energy resolution of these detectors, together with the very low background spectrum of 
dedicated counting \mbox{setups}. Radioactive contaminants are identified by the gamma lines associated with their decay \cite{ref::heusser}.
The spectrometers used for the 
measurements presented here are installed underground at LNGS, at a depth of 3100~meters water equivalent relative to a flat overburden, where the muon flux is reduced
by a factor $\sim 10^{6}$ with respect to above ground. This also reduces the 
neutron flux by several orders of magnitude since the hadronic showers from cosmic rays are completely 
blocked and muon induced neutrons are strongly suppressed. The remaining neutron flux is mainly due
to fission and $(\alpha,n)$ reactions. A ventilation system constantly brings fresh air from the outside to
the laboratory, leading to a rather constant $^{222}$Rn concentration of $\sim$30 Bq/m$^3$ at the location
of the HPGe spectrometers. For samples with mass of the order 1 kg or more, this technique
allows to reach sensitivities of $< 1$ mBq/kg, corresponding to $\sim 10^{-10}$ g $\times$ g$^{-1}$ levels of U and Th
contamination.

In this section we briefly describe the facilities used for this study and how the measurements were analyzed. 

\paragraph*{Gator Facility}

Gator \cite{gator_paper} is a High Purity p-type coaxial germanium detector of 2.2~kg sensitive
mass, a relative efficiency of 100.5\%\footnote{The efficiency is defined relative to a 7.62 cm
$\times$ 7.62 cm NaI(Tl) crystal, for the 1.33MeV $^{60}$Co photopeak, at a source-detector
distance of 25cm \cite{knoll}.} and a measured resolution of $\sim$ 3~keV FWHM at 1332~keV.
In order to ensure high detection sensitivities,
the detector and its shield have been constructed using materials selected for their 
extremely low intrinsic radioactive contaminations. 
The shield consists of 20 cm of lead (15 cm with a $^{210}$Pb activity of 75 Bq/kg 5 cm with 3 Bq/kg),
and 5 cm of oxygen-free high-conductivity (OFHC) copper. It fulfills the
crucial requirements of having sufficient sample capacity and low intrinsic background,
and uses nitrogen purging against radon and its progenies.
The inner dimensions of the sample cavity are $25 \times 25 \times 33$~cm$^3$, with the 
detector reaching into the cavity. The total available volume is $\sim$19~liters.
An example measurement, showing the different gamma lines for the spectrum of
a sample measured with Gator, is shown in Fig.~\ref{fig::r8520}.

\paragraph*{LNGS Counting Facility}
The low background germanium counting facility of LNGS \cite{Arpesella} 
is equipped with several high-purity
coaxial germanium detectors with large sensitive volumes, low intrinsic background and
high energy resolution ($<2.6$~keV FWHM at 1332~keV). They include GeMPI-I and GeMPI-II \cite{GeMPI}, 
the most sensitive screening facilities worldwide. For all detectors only 
selected materials with lowest intrinsic radioactivity have been used.
They are shielded with 25 cm thick layers of lead with an activity of 20 Bq/kg in $^{210}$Pb
and 10 cm of OFHC copper, except for the GeMPI detectors which have a shield made of 20 cm of lead with decreasing
$^{210}$Pb concentration (5 Bq/kg for the innermost layer) and with an inner layer of 5 cm of OFHC copper.
The sample cavities have volumes which range
from one liter up to 15 liters (GeMPI-I/II) which are constantly purged with
nitrogen gas.

\paragraph*{Measurement}
In order to increase the sensitivity of a $\gamma$-counting measurement,
it is beneficial to use massive
samples. Since placing the samples very close to the detector also
increases the efficiency, the best configuration is to place
the sample on top of the germanium crystal.
For large or heavy samples, however, this requirement cannot always be
fulfilled and the sample has to be placed around the detector. 

Before the measurement, the samples are properly cleaned from surface contaminations 
in an ultrasonic bath of pure ethanol ($>$ 98\%). 
Afterwards they have been stored in an environment purged with pure boil-off 
N$_2$ for several days in order to let $^{222}$Rn and its progenies decay or diffuse out.
Once the samples are inserted into the spectrometers, the $\gamma$-ray spectra are acquired 
automatically using standard multi-channel analyzers. The measurement time varied according to 
the sample. Typically it was in the range of days to several weeks.
Data analysis has been performed off-line after the measurement.

As the geometrical arrangement of every sample is unique, the detection efficiency 
has been calculated for every measurement using Monte Carlo simulations. For
Gator, the GEANT4 toolkit \cite{GEANT4}
is used, while the LNGS facility detectors employ simulations based on GEANT4 and
GEANT3.21. In all cases, the whole
sample and detector geometry is coded. Standard sources of known activity are regularly 
used to test the reliability of this method of efficiency estimation, giving excellent
agreement \cite{gator_paper}.

\paragraph*{Analysis}\label{sec:Analysis}
To determine the concentration of a specific radioactive decay chain, the
characteristic $\gamma$-lines of the chains are considered: \\
-- $^{232}$Th chain: $^{228}$Ac, $^{212}$Pb, $^{212}$Bi and $^{208}$Tl. \\
-- $^{238}$U chain: $^{234m}$Pa, $^{234}$Th, $^{214}$Pb and $^{214}$Bi. \\
-- $^{235}$U chain: $^{235}$U. \\
In some cases the activity of $^{235}$U is deduced from that of $^{238}$U by
taking into account their relative abundance in natural uranium (0.70\%
of $^{235}$U and 99.27\% of $^{238}$U).
The activities of the different decay chains are calculated taking into account the mass of the
sample, the measurement time, the efficiency from the Monte Carlo simulation and the
number of counts in the characteristic $\gamma$-line. The background spectrum, which is measured
regularly and the
Compton continuum from other lines are subtracted. More details on the analysis
can be found in \cite{gator_paper}.
Upper limits are calculated using the method introduced in \cite{ref::hurtgen} for Gator
and in \cite{Heisel} for the LNGS screening facility.

Secular equilibrium can be verified for $^{232}$Th because of the relatively high branching
ratios of the $\gamma$-lines emitted in the initial part of the chain. For most of the screened samples
this is not possible for $^{238}$U. However, upper limits obtained from
the $\gamma$-lines of the daughters $^{234m}$Pa and $^{234}$Th do not exclude
the secular equilibrium hypothesis. The only samples for which
a break in the secular equilibrium of $^{238}$U has been detected are
the field shaping resistors and the ceramic feedthrough (entries~44 and~46 in 
Table~\ref{tab:screen}): the radioactivity of the
different parts of the decay chain are given explicitely there.

All other relevant radioactive nuclides contributing to the $\gamma$-ray spectra 
(such as $^{40}$K, $^{60}$Co, $^{137}$Cs, etc.) are analyzed 
using their most prominent $\gamma$-lines 
at specific energies and branching ratios.

\section{Inductively Coupled Plasma Mass Spectrometry}\label{icpsm}

For a few plastic samples, with a total mass too small to yield reasonable sensitivity with
the HPGe detectors, inductively coupled Plasma-Mass Spectrometry (ICP-MS) has been used to determine
the intrinsic radioactivity. 
The ICP-MS~7500a from {\it Agilent Technologies} employed here is 
a quadrupolar mass spectrometer using an argon plasma torch (ICP-torch) to ionize the sample.

Before the procedure, the samples are cleaned and prepared as described in Section \ref{sec::gammaray}. 
They are then ashed at 600$^\circ$C, using the dry ashing procedure
described in \cite{Gaines2003} and dissolved in a 10\% ultra pure
nitric acid solution, which is nebulized in a spray chamber where it forms an aerosol. 
The aerosol is atomized and ionized by the ICP-torch, producing a cloud of
positively charged ions. The ions are extracted from the plasma
into an ultra high vacuum system containing a quadrupole analyzer,
where they are separated according to their mass-to-charge ratio $q$.
The count rate obtained for a particular $q$ 
is compared with a calibration curve to determine the concentration
of the elements in the sample. The sensitivity of the Mass Spectrometer
used for the measurements presented here is of the order of $10^{-11}$ g $\times$ g$^{-1}$ for Uranium
and Thorium (corresponding to $\sim$ 0.1 mBq/kg) and $10^{-7}$ g $\times$ g$^{-1}$ for K
(corresponding to $\sim$ 1 mBq/kg) \cite{Nisi:2009}.

For the samples screened with the ICP-MS secular equilibrium is assumed.

\section{Results and Discussion}\label{sec::results}

The main results from the screening campaign are presented in Table \ref{tab:screen}. 
It lists the material, the supplier and its use in the XENON100 experiment (see also 
Section~\ref{sec::xe100}). It 
provides details on the measurement (detector, measurement time and sample mass) 
and gives the measured radioactive contaminations or, where no spectral lines have been observed,
upper limits at 95\% confidence level. In this Section, the most relevant results are discussed.

\subsection{Metal Samples}

Lead is used for the XENON100 passive shield. Material from
two different suppliers was screened (entries 1-4 in Table~\ref{tab:screen}). The lead from
{\it Plombum FL} is a standard lead while the one from {\it Founderies de Gentilly}
has a low contamination of $^{210}$Pb. This was confirmed in both HPGe facilities, 
however, measured with different sample masses.
When evident lines are detected, the results are in agreement, 
when only upper limits can be given, the limits
from the measurement with more mass are considerably lower.
The $^{210}$Pb activities are $(530 \pm 70)$~Bq/kg and $(26 \pm 6)$~Bq/kg for lead with 
high and low $^{210}$Pb contamination, respectively.

One of the most radiopure metals which is commonly available is OFHC copper. 
This makes it a very interesting
candidate to be used for the innermost structural parts of low background detectors and it
is a very pure material for shielding. Large samples of
copper typically give null results, even in the most sensitive spectrometers available
\cite{GeMPI}. The copper used in the XENON100 shield (entry~5 in Table ~\ref{tab:screen}) is the same as used in
the shield of the Gator spectrometer. From the detailed background
model of this facility,
copper activities of $(75\pm14)$~$\mu$Bq/kg for
$^{226}$Ra and $(21\pm7)$~$\mu$Bq/kg for $^{232}$Th have been determined \cite{gator_paper}. 
Of all samples presented in this paper, where a detection can be claimed, these are the ones 
with the lowest radioactivity contaminations.

The second copper sample in the Table (entry~6) is from the same provider
({\it Norddeutsche Affinerie AG}) and its radioactive contamination with
$^{226}$Ra and $^{232}$Th is below the sensitivity of
the spectrometer. At the time of the screening, it had, however, a non negligible concentration of $^{60}$Co
since it had been stored at the Earth's surface for several months where it was activated by cosmic rays.  

Several samples of austenitic
\footnote{This steel has the reference \it{X 6 CrNiMoTi 17 12 2} in the European
EURONOI standard and \it{316L} in the US AISI standard.}
stainless steel type 1.4571, all supplied by {\it NIRONIT Edelstahlhandel GmbH
\& Co.\ KG}, were screened and confirm the results published in~\cite{GERDASS}
(entries 7--10). The sample sheets had a different thickness. 
The content of $^{226}$Ra and $^{232}$Th in all samples is below 4~mBq/kg. Lines were 
detected only in the 3~mm and 25~mm thick samples (entries 9 and 10,
respectively) which had the largest sample mass. This is in agreement with~\cite{GERDASS}.
The contamination of $^{60}$Co is even lower than the measurements presented there. The 
values range from $(1.4 \pm 0.3)$~mBq/kg to $(13 \pm 1)$~mBq/kg, which are the lowest
values ever published for stainless steel. Radioactivity generated by cosmic ray activation during 
surface exposure was detected in all the samples, with the
longest lived isotope being $^{54}$Mn ($\tau_{1/2}=312$~days) which was found with
activities ranging from $(1.7\pm0.4)$~mBq/kg in the 25~mm sample to $(0.5\pm0.2)$~mBq/kg in
the 2.5~mm sample.

\subsection{Plastic Samples}

In direct Dark Matter detection experiments, 
polyethylene (entries 12--14) is commonly used as neutron moderator in radiation shields 
surrounding the experiments.
The polyethylene from the XENON100 shield was measured in different quantities and for different
measurement times. All measurements are compatible with each other.
$^{226}$Ra, $^{40}$K and $^{137}$Cs are detected when a large sample
(8.44~kg) is measured for a relatively long time (28.9~days), but the respective
contaminations are all below 0.7~mBq/kg. However, following these results it was
decided to add a 5~cm layer of OFHC copper to the XENON100 shield, inside the Polyethylene, in order to 
suppress this $\gamma$-background.

Polytetrafluoroethylene (PTFE, Teflon) is a material widely used in liquid xenon
applications because of its physical, mechanical, dielectric and optical properties.
It withstands liquid xenon (LXe) temperatures ($-95^\circ$C) and is
a good insulator with a dielectric constant very similar to LXe ($\epsilon_r=2$). 
It also is an excellent reflector for VUV~light 
at the xenon scintillation wavelength $\lambda=178$~nm~\cite{Yamashita:2004}. 

A possible drawback of using PTFE in Dark Matter experiments is that fluorine $^{19}$F, the
main component of the PTFE, has a high cross section for
$(\alpha,n)$ reactions ($\sim\!200$~mb for \mbox{$E_{\alpha}=5.5$~MeV}
\cite{Norman:1984}). The $\alpha$-particles for these reactions would come 
from the $^{226}$Ra and $^{232}$Th chains from intrinsic contaminations. Therefore, a number of measurements were
performed with the PTFE used for the construction of
the XENON100~TPC (entries 15--16 in Table~\ref{tab:screen}). No evidence 
for radioactive contaminants in the PTFE has been found within the sensitivity of the used
spectrometer and only upper limits could be derived. In particular the contamination with
$\alpha$-emitters is found to be $<$0.1~mBq/kg for $^{232}$Th and $<$0.06~mBq/kg
for $^{226}$Ra. Other samples of PTFE from different suppliers were measured (with a lower sensitivity)
and found to have radioactive contaminations below 2~mBq/kg (entries 17--19 in Table~\ref{tab:screen}).

\onecolumn
\input{xe100_screening_table_split}
\twocolumn

Because of their small mass, two plastic samples were screened using the ICP-MS technique
described in Sect.~\ref{icpsm}: a polymethyl methacrylate (PMMA) optical fiber to guide blue light from a LED into
XENON100 for photosensor calibration (entry~41) and a thin PTFE sheet used as a light reflector (entry~17).
The PTFE sheet has a radioactive contamination well within the XENON100 radioactivity requirements.
Although the optical fiber exhibits a higher radioactivity than other plastic samples its contribution to the
XENON100 background is negligible with respect to other samples, given its very small mass (10~g total).

\subsection{R8520-06 Hamamatsu PMTs}

Among the most important samples presented in this paper are the R8520-06-Al square
$1'' \times 1''$ photomultiplier tubes (PMTs) from {\it Hamamatsu}. They are the light sensors
chosen for XENON100 and therefore a central part of the detector. Given their
proximity to the target volume and their nature of being a composite, fully assembled object, they
are one of the dominating background sources. In order to determine the 
contribution of the PMTs to the overall background, 
and to reject possible ``hot'' tubes, a large fraction of the PMTs installed in the detector
has been screened (entries 20--36 in Table~\ref{tab:screen}) in the various facilities.
The average contamination of $^{238}$U, $^{232}$Th, $^{40}$K and $^{60}$Co 
for the fraction of PMTs installed in XENON100 is
$(0.25 \pm 0.04)$, $(0.5 \pm 0.1)$, $(8.1 \pm 0.9)$ and $(0.75 \pm 0.08)$ mBq/PMT, respectively.

\begin{table*}[t!]
\caption{\label{tab:pmtparts} Summary of the measured activities of the individual parts of the {\it Hamamatsu} R8520-06 PMT. The last row gives the expected values for a single PMT based on the measurements (1-6) taking into account the mass model, given in the second column of the Table.}
\footnotesize
\begin{center}
\begin{aenumerate}
\begin{tabular}{p{4.4cm}p{0.9cm}p{1.0cm}p{1.0cm}p{1.0cm}p{1.1cm}p{1.0cm}p{1.0cm}p{1.0cm}p{1.4cm}}
\hline
\hline
\multirow{2}{*}{\bf{PMT Component}}										&{\bf{Mass}}		&\multicolumn{2}{c}{\bf{$^{238}$U}}		&\multicolumn{2}{c}{\bf{ $^{232}$Th }}	&\multicolumn{2}{c}{\bf{$^{40}$K }}		&\multicolumn{2}{c}{\bf{$^{60}$Co}} \\
														& {\bf g}										&\bf{mBq/kg}		&\bf{mBq/PMT}			& \bf{mBq/kg}		&\bf{mBq/PMT}			& \bf{mBq/kg}		&\bf{mBq/PMT}			& \bf{mBq/kg}		&\bf{mBq/PMT}			\\	
\hline
\hline
\aitem Kovar Metal: main metal package									&	13						&	19(7)		&	0.25(9)		&	$<13$		&	$<0.17$	&	90(10)	&	1.2(1)		&	40(20)	&	0.5(3)\\
\aitem Borosilicate glass: used in stem									&	1						&	970(20)		&	0.97(2)		&	340(20)		&	0.34(2)		&	2300(200)&	2.3(2)		&	$<10$		&	$<0.01$	\\
\aitem Ceramic: spacer between electrodes 								&	0.04						&	780(20)		&	0.031(1)		&	260(20)		&	0.010(1)	&	800(100)	&	0.032(4)		&	$<12$		&	$<4.8\times10^{-4}$\\
\aitem Aluminum: sealing between \\ \mbox{\ \ \ \ quartz window and metal package}			&	0.1						&	17(8)		&	0.0017(8)		&	370(20)		&	0.37(2)		&	5(2)  		&	0.0005(2)		&	$<$ 0.27    	&	$<2.7\times10^{-5}$ \\
\aitem Stainless Steel: electrodes  										&	7						&	19(7) 		&	0.13(5)		&	18(8)		&	0.13(6)	&0.15(2)		&	1.1(1)		&	12(5)		&	0.08(3) \\
\aitem Glass (synthetic silica): window 									&	2						&	$<$0.5		&	$<0.001$		&	$<1.8$		&	$<0.0036$&18(3)		&	0.036(6)		&	$<0.1$		&	$<2\times10^{-4}$\\
\hline
R8520-06-AL (total) 													&	23.14					& 				&1.4(2)		&				&0.85(7)		&				&4.6(2)		&				&0.6(3) \\
\hline
\hline
\end{tabular}
\end{aenumerate}
\end{center}
\normalsize
\end{table*}

In order to fulfill the design requirements of XENON100,
the manufacturer of the PMTs has assured that the components used for the production of the
various PMT batches always have the same controlled origin. Nevertheless, some of the processes or
working conditions in the production of the components might differ. Therefore,
the PMTs have been subdivided in groups of typically 10-20 PMTs with tubes from the same  
production batch. The different groups have been screened independently to
cross check their intrinsic radioactive contamination. Some batches   
have been discarded because of their increased level 
of radioactivity (entries 22, 24 and 25), which does not fulfill the design requirement. 
The PMTs in entry 25 were a subsample of devices used in XENON10. They have not been
re-used in XENON100 because of their high $^{60}$Co and $^{137}$Cs content.
All other screened PMT batches met the radioactivity requirements and were installed in the detector.

In order to have a detailed characterization of this PMT model and the possibility 
to improve the radioactivity,
{\it Hamamatsu} has provided large amounts ($> 100$ grams)
of each component of the R8520 PMT used in XENON10 \cite{Xe10_prl_SI}.
These have been screened individually and the results are given in Table~\ref{tab:pmtparts}.

According to the PMT mass model provided by {\it Hamamatsu}, 
(see second column of Table~\ref{tab:pmtparts}), the main components responsible
for the overall PMT radioactivity have been identified: these are the metal package and 
the stem pins (made of Kovar metal which has a small thermal expansion coefficient), 
the borosilicate glass in the stem and the stainless steel electrodes.
The table shows the contributions of the single components to the
overall PMT radioactivity. Knowing the 
constituents dominating the overall radioactivity allows to select other materials, 
or materials processed in a different way, to use them in the development of new 
photosensors. This is important for the next generation of rare event search experiments where the
background requirements are even more demanding. 

The predicted activities for a single PMT, based on these measurements of the individual components,
are given in the last row of Table~\ref{tab:pmtparts}. The numbers have to be compared with entries
20--36 of Table~\ref{tab:screen}. For $^{238}$U and $^{232}$Th the predictions are higher than
the average of the PMTs used in XENON100, but somewhat lower in $^{60}$Co
and $^{40}$K. However, since the individual parts were not from the batches used in XENON100,
and the disagreement is only a little larger than the uncertainties, this is an acceptable
result.

\subsection{Other samples}

New photosensors with a larger active area (entries 37--39 in Table \ref{tab:screen})
have also been  examined as they are interesting for the next generation LXe detectors such as XENON1T \cite{QUPIDs}.

Special effort went into the design of the voltage divider
circuit (bases) for the R8520-06 PMTs used in XENON100. The components 
(surface mount resistors and capacitors on a Cirlex substrate)
have been reduced in mass and number and have been selected for lowest possible
radioactive contaminations (entry 40 in Table~\ref{tab:screen}). 
However, the contribution of the bases to the total background at low energies is still 10\% of that from the PMTs \cite{Xe100BG}.
The cables for the PMT signal (RG174, RG174, with the outer plastic insulation removed)
and high voltage (silver coated, kapton insulated) have also been measured (entries 42 and 43). 
Given the rather small amount used in XENON100, their impact on the overall background is negligible.

Finally, another sample screened within the scope of this paper is the concrete
used to build the underground laboratory. Two samples were taken from the wall and
from the floor surrounding the XENON100 installation. The precise knowledge of the concrete's 
radioactive contaminations allows to calculate the expected neutron flux from
$(\alpha,n)$ and spontaneous fission reactions in the material. The results measured with
the Gator facility (entries 48, 49) agree with measurements of the
background $\gamma$-flux at the same location~\cite{ref::lngs_bg}.

\subsection{Location of the screened components in XENON100}\label{sec::xe100}

This section summarizes the location of the screened materials, as listed in Table \ref{tab:screen},
in XENON100. The detector is a
position-sensitive TPC using liquid xenon (LXe) as target material. The light
signals are detected by two arrays (on top and bottom of the detector) of {\it Hamamatsu} R8520-06 PMTs (entries
20--36). The LXe volume outside the TPC is also instrumented with PMTs in order to act as an active veto. 
All PMTs are mounted on voltage dividers (entry 40) and connected
with coaxial (signal, entry 42) and single wire cables (high voltage, entry 43) to the feedthroughs (entry 46). 
These are placed outside the shield because of their higher radioactivity level.
The sensitive target can be ``fiducialized'' to suppress the background by only keeping the inner core
of the detector for the science analysis, exploiting the good self shielding properties of the LXe.

In order to suppress the $\gamma$-background from the laboratory environment (mainly from concrete \cite{ref::lngs_bg}, entries 48 and 49), the detector is placed inside a passive shield. It is made from lead (15~cm ``normal'' lead, entries 1 and 2,
followed by 5~cm lead low in $^{210}$Pb, entries 3 and 4), 20~cm polyethylene (entries 12--14) and 5~cm OFHC copper (entry 5). The whole shield sits on a 25~cm slab of polyethylene and 
is additionally shielded against neutrons with 20~cm water on 3~sides and on the top. 

The detector is installed in a double-walled cryostat made from low radioactivity stainless steel (entries 7-10).
The LXe target volume is enclosed by a PTFE cylinder (entry 16) of $\sim$15~cm radius and $\sim$30~cm height.
Copper wires, wound around the TPC and connected with resistors (entry 44), 
ensure electric field homogeneity. The PTFE panels
are stabilized using rings of OFHC copper (entry 6). The same copper supports the PMTs on the bottom of the
detector and all PMTs in the active LXe veto. The PMTs above the target are resting in a
PTFE structure (entry 16) and fixed by smaller copper rings (entry 6). 
Everything is held together with stainless steel
screws (entry 11). In order to improve the light collection in the active LXe veto volume
the cryostat wall is covered by a thin PTFE sheet (entry 17).

A complete study of the XENON100 background has been published in \cite{Xe100BG}. It uses the screening
results reported here as input values.

\section{Summary}

The materials used to construct experiments for rare event searches have to be 
selected  in order to achieve the lowest possible radioactive background. 
This paper presents the results from the radioactivity screening campaign for XENON100, 
which aims to directly detect WIMP Dark Matter.

More than 20~different materials have been examined, mostly using low
background HPGe detectors, but also applying mass spectrometry. In
many cases, several batches of samples have been screened to check systematics or 
because the material properties were slighly different (different production batches, thickness, etc.). 
All results are given in Table~\ref{tab:screen}, which might be very useful for
other experiments searching for rare events. For this reason, we also provide the supplier for all samples. 

These results have been used in a study to predict the electromagnetic background of XENON100 \cite{Xe100BG}.
By comparing the measured energy spectrum between threshold and 2700~keV to a detailed Monte Carlo analysis, 
it is verified that the background design goal of $< 10^{-2}$~events~keV$^{-1}$~kg$^{-1}$~day$^{-1}$ has
indeed been reached.  
Of all running direct Dark Matter detection experiments, XENON100 has the lowest electromagnetic
background. This has already
allowed to set a competitive limit on spin-independent WIMP-nucleon interactions with only a few
days of measuring time \cite{xe100_prl_first, xe100::pl}.

A detailed study of the intrinsic radioactive contamination of 
the {\it Hamamatsu} R8520-06 PMT, the light sensor used in XENON, has also been presented. The screening of 
individual PMT components provides information about which parts have to be modified in 
order to further decrease the radioactivity level. Almost equally important are the results
obtained for the PTFE used for the XENON100 TPC, which is one of the 
purest PTFE samples ever reported in the literature.

\section*{Acknowledgments}
This work has been supported by the National Science Foundation Grants No.~PHY-03-02646 and
PHY-04-00596, the Department of Energy under Contract No.~DE-FG02-91ER40688, the CAREER
Grant No.~PHY-0542066, the Swiss National Foundation Grant No.~20-118119 and No.~20-126993, the Volkswagen
Foundation and the FCT Grant No.~PTDC/FIS/100474/2008.

We would like to thank the Max Planck Institut f\"ur Kernphysik, Heidelberg, for giving us screening time on the GeMPI detectors.

We thank Giuseppina Mosca from the LNGS chemistry laboratory for the
assistance in cleaning the samples. 


\end{document}

%% file: xe100_screening_table_split.tex
%
%

\footnotesize
\begin{landscape}
\setlength{\LTcapwidth}{8.8in}
\begin{zenumerate}
\begin{longtable}{p{0.1in}p{0.9in}p{.57in}p{0.8in}p{.45in}p{.45in}p{.35in}p{.48in}p{.4in}p{.4in}p{.4in}p{.4in}p{.4in}p{.4in}p{.45in}p{.45in}}
\caption{Screening Results. See text for discussion.}
\label{tab:screen}\\
\hline
\hline
	&	\bf{Material} 	&	\bf{Supplier}	&	\bf{Use}	&	\bf{Detector}		&	\bf{Time [d]} 	&	\bf{Amount}		&	\bf{Unit}		&	\bf{$^{228}$Ra}		&	\bf{$^{228}$Th}&	\bf{$^{238}$U}	&	\bf{$^{226}$Ra}			&	\bf{$^{235}$U}		&	\bf{$^{40}$K}		&	\bf{$^{137}$Cs}		&	\bf{$^{60}$Co}	\\ \hline \hline
\endfirsthead
\caption{Screening Results (continued).}\\
\hline 
\hline 
	&	\bf{Material} 	&	\bf{Supplier}	&	\bf{Use}	&	\bf{Detector}		&	\bf{Time [d]} 	&	\bf{Amount}		&	\bf{Unit}		&	\bf{$^{228}$Ra}		&	\bf{$^{228}$Th}&	\bf{$^{238}$U}	&	\bf{$^{226}$Ra}			&	\bf{$^{235}$U}		&	\bf{$^{40}$K}		&	\bf{$^{137}$Cs}		&	\bf{$^{60}$Co}	\\ \hline \hline
\endhead

\hline \hline \endfoot
\hline \hline \endlastfoot
	&	\bf{Metal}	&		&		&		&		&		&		&		&		&		&		&		&		\\ 
\zitem	&	Lead	
		&	Plombum
		&	Outer Pb shield
		&	Gator 		
		&	18.4		
		&	2.27 kg 	
		&	mBq/kg
		&	$<6.9$		
		&	$< 0.52$
		&	$<260$
		&	$<4.2$
		&	$<12$
		&	14(3)
		&	$<0.81$
		&	$< 0.11$	\\
\zitem	&	Lead
		&	Plombum	
		&	Outer Pb shield	
		&	LNGS 				
		&	14.5		
		&	44 kg 	
		&	mBq/kg
		&	$< 6.6 $
		&	$< 1.6 $
		&	$<130$
		&	$< 5.7$
		&	$< 51$
		&	14(6)
		&	$< 2.1$
		&	$< 1.1$	\\
\zitem	&	Lead
		&	Foundaries \mbox{de Gentilly}	
		&	Inner Pb shield	
		&	Gator
		&	17.8		
		&	2.27 kg 	
		&	mBq/kg				
		&	$< 0.66$
		&	$< 0.42$
		&	$<24$	
		&	$< 0.71 $	
		&	$<1.8$
		&	$< 1.46$
		&	0.63(6)
		&	$< 0.11$	\\
\zitem  	&	Lead
		&	Foundaries \mbox{de Gentilly}	
		&	Inner Pb Shield	
		&	LNGS				
		&	18.7
		&	44 kg
		&	mBq/kg	
		&	$< 3.9 $
		&	$< 4.3 $
		&	$<33$	
		&	$< 6.8$
		&	$< 20$
		&	$< 28 $
		&	$< 0.85$
		&	$< 0.19 $	\\
\zitem	&	Copper
		&	Norddeutsche \mbox{Affinerie}
		&	Shield
		&	Gator
		&	51.4
		&	512 kg
		&	$\mu$Bq/kg
		&	21(7)
		&	21(7)
		&	70(20)
		&	70(20)
		&	3.4
		&	23(6)
		&	
		&	2(1)	\\
\zitem	&	Copper
		&	Norddeutsche \mbox{Affinerie}
		&	TPC
		&	Gator
		&	20.3
		&	18.1 kg
		&	mBq/kg
		&	$<0.37 $
		&	$<0.33 $
		&	$<11$
		&	$< 0.37$
		&	$<0.47$
		&	$<1.3   $
		&	 $<0.14$
		&	0.24(6)     	\\
\zitem	&	\mbox{Stainless Steel} \mbox{316Ti (1.5 mm)}
		&	NIRONIT
		&	Cryostat wall
		&	LNGS
		&	6.87
		&	1.2 kg
		&	mBq/kg
		&	$< 2.4 $
		&	$< 1.0 $
		&	$<130$
		&	$< 1.9$
		&	$< 2.0 $
		&	10(4)
		&	$< 0.9 $
		&	8.5(9) 	\\
\zitem	&	\mbox{Stainless Steel} \mbox{316Ti (2.5 mm)}
		&	NIRONIT
		&	Cryostat bottom
		&	LNGS
		&	20.6
		&	1.97 kg
		&	mBq/kg
		&	$< 3.1 $
		&	$< 1.5 $
		&	$<42$
		&	$< 2.7 $
		&	$< 1.4 $
		&	$< 12 $
		&	$< 0.88 $
		&	13(1) 	\\
\zitem	&	\mbox{Stainless Steel} \mbox{316Ti (3.0 mm)}
		&	NIRONIT
		&	Grid frame
		&	Gator
		&	6.76
		&	6.6 kg
		&	mBq/kg
		&	$<4.1$
		&	$<1.8$
		&	$<130$
		&	3.6(8)
		&	$<5.8$
		&	$< 5.7 $
		&	$<1.1$
		&	7(1)	 	\\
\zitem	&	\mbox{Stainless Steel} \mbox{316Ti (25 mm)}
		&	NIRONIT
		&	\mbox{Top~flange/} \mbox{Support bars}
		&	LNGS
		&	5.58
		&	1.52 kg
		&	mBq/kg
		&	$<0.92$
		&	2.9(7)
		&	$< 20 $
		&	$<1.3$
		&	$<1.3 $
		&	$<7.1 $
		&	$<0.82 $
		&	1.4(3) 	\\
\zitem	&	Screws 2-56 7/16"
		&	McMaster
		&	\mbox{Standard screw}
		&	Gator
		&	12.1
		&	0.27kg
		&	mBq/kg
		&	24(5)
		&	$<21$
		&	$<550$
		&	$<13$
		&	$<25$
		&	$<$ 47
		&	$<5.1$
		&	6(2)          \\
\\ \hline
	&	\bf{Plastic}	&		&		&		&		&		&		&		&		&		&		&		&		\\
\zitem	&	Polyethylene
		&	in2plastic
		&	Shield wall
		&	Gator
		&	5.85
		&	2.76 kg
		&	mBq/kg
		&	$< 5.4$
		&	$< 3.7$
		&	$<170$
		&	$< 5.1$
		&	$<7.6$
		&	$< 14 $
		&	 $<1.7$
		&	$< 1.4$	\\
\zitem	&	Polyethylene
		&	in2plastic
		&	Shield door
		&	Gator
		&	3.12
		&	3.1 kg
		&	mBq/kg
		&	$< 4.3$
		&	$< 5.8$
		&	$<220$
		&	$< 6.5 $
		&	$<9.9$
		&	$< 13$
		&	$<2.1$
		&	$< 1.7 $  	\\
\zitem	&	Polyethylene
		&	in2plastic
		&	Shield wall/door
		&	LNGS
		&	28.9
		&	8.44 kg
		&	mBq/kg
		&	$< 0.094$
		&	$<0.14$
		&	$<3.8$
		&	0.23(5)
		&	 $< 0.37 $
		&	0.7(4)
		&	 0.06(3)
		&			\\
\zitem	&	PTFE
		&	Maagtechnic
		&	TPC
		&	Gator
		&	14.35
		&	13.5 kg
		&	mBq/kg
		&	$< 0.39$
		&	$< 0.16$
		&	$<6.2$
		&	$< 0.31$
		&	$<0.28$
		&	$< 2.25$
		&	$<0.13$
		&	$< 0.11$	\\
\zitem	&	PTFE
		&	Maagtechnic
		&	TPC
		&	Gator
		&	47.4
		&	23.5 kg
		&	mBq/kg
		&	$<0.16$
		&	$<0.10$
		&	$<3.0$
		&	$<0.06$
		&	$<0.13$
		&	$<0.75$
		&	$<0.07$
		&	$< 0.03$	\\
\zitem	&	PTFE
		&	McMaster
		&	Veto reflector
		&	ICP-MS
		&
		&	5.1 g
		&	mBq/kg
		&	0.5(1)
		&	0.5(1)
		&	0.25(5)
		&	0.25(5)
		&	0.011(2)
		&	$< 3.1 $
		&
		&		\\
\zitem	&	PTFE
		&	McMaster
		&	XENON10 TPC	
		&	LNGS
		&	10.1
		&	0.23 kg
		&	mBq/kg
		&	$<1.8$
		&	$<2.3$
		&	$<36$
		&	$<1.1$
		&	$<1.4$
		&	$<7.6 $
		&	$< 0.44$
		&		\\
\zitem	&	PTFE
		&	APT
		&	Not used
		&	LNGS
		&	23.5
		&	6.54 kg
		&	mBq/kg
		&	$< 0.15 $
		&	$< 0.13 $
		&	$<12$
		&	$< 0.16$
		&	$< 0.59$
		&	3(1)
		&	$< 0.11$
		&	0.15(7)	\\ \hline 
\newpage
	&	\bf{Light Sensors}	&		&		&		&		&		&		&		&		&		&		&		&		\\
\zitem	&	R8520 - Batch 1
		&	Hamamatsu
		&	Top array, veto
		&	LNGS
		&	11.6
		&	7 pc
		&	mBq/PMT
		&	$<0.32$
		&	0.19(3)
		&	$<5.3$
		&	0.15(3)
		&	$< 0.13$
		&	10(1)
		&	$< 0.05$
		&	0.56(5) 	\\
\zitem	&	R8520 - Batch 2
		&	Hamamatsu
		&	Bottom array
		&	LNGS
		&	21.6
		&	7 pc
		&	mBq/PMT
		&	$<0.22$
		&	0.16(4)
		&	$<5.2$
		&	0.21(3)
		&	0.10(4)
		&	9(1)
		&	$< 0.05$
		&	0.59(5) 	\\
\zitem	&	R8520 - Batch 3
		&	Hamamatsu
		&	Not used
		&	LNGS
		&	6.65
		&	1 pc
		&	mBq/PMT
		&	$<2.1$
		&	1.9(5)
		&	$<75$
		&	 0.6(1)
		&	$< 0.70$
		&	 120(10)
		&	$< 0.51$
		&	4.5(5)	\\
\zitem	&	R8520 - Batch 4
		&	Hamamatsu
		&	Top array
		&	LNGS
		&	18.4
		&	7 pc
		&	mBq/PMT
		&	$<0.23$
		&	0.19(3)
		&	$<2.6$
		&	0.25(4)
		&	$< 0.08$
		&	7(1)
		&	$< 0.05$
		&	 0.67(6)	\\
\zitem	&	R8520 - Batch 5
		&	Hamamatsu
		&	Not used
		&	LNGS
		&	12.8
		&	4 pc
		&	mBq/PMT
		&	$<0.49$
		&	0.38(8)
		&	$<9.9$
		&	0.25(4)
		&	$< 0.08$
		&	7(1)
		&	$< 0.05$
		&	 0.67(6)	\\
\zitem	&	R8520 - Batch 6
		&	Hamamatsu
		&	Not used
		&	LNGS
		&	10.4
		&	5 pc
		&	mBq/PMT
		&	$<0.42$
		&	0.38(8)
		&	$<9.9$
		&	0.39(8)
		&	$< 0.24$
		&	12(2)
		&	17(6)
		&	 2.7(2) 	\\
\zitem	&	R8520 - Batch 7
		&	Hamamatsu
		&	Top/bottom array, veto
		&	LNGS
		&	24.4
		&	39 pc
		&	mBq/PMT
		&	0.087(3)
		&	0.11(1)
		&	$<4.7$
		&	0.12(1)
		&	0.04(1)
		&	6.9(7)
		&	0.027(7)
		&	1.5(1)		\\
\zitem	&	R8520 - Batch 8
		&	Hamamatsu
		&	Top array, veto
		&	LNGS
		&	11.9
		&	48 pc
		&	mBq/PMT
		&	$<0.11$
		&	0.11(1)
		&	$<1.4$
		&	0.12(1)
		&	0.04(1)
		&	7.7(8)
		&	$< 0.020 $
		&	0.56(4)		\\
\zitem	&	R8520 - Batch 9
		&	Hamamatsu
		&	Bottom array
		&	LNGS
		&	4.7
		&	23 pc
		&	mBq/PMT
		&	0.5(2)
		&	0.22(4)
		&	$<2.7$
		&	0.16(5)
		&	$< 0.073$
		&	14(2)
		&	 $< 0.021$
		&	0.73(7)		\\
\zitem	&	R8520 - Batch 10
		&	Hamamatsu
		&	Bottom array
		&	Gator
		&	5.5
		&	22 pc
		&	mBq/PMT
		&	$<0.59$
		&	0.3(1)
		&	$<15$
		&	$<0.28$
		&	$<0.67$
		&	11(1)
		&	$<0.1$
		&	0.53(6)		\\
\zitem	&	R8520 - Batch 11
		&	Hamamatsu
		&	Veto
		&	LNGS
		&	17.1
		&	15 pc
		&	mBq/PMT
		&	0.19(3)
		&	0.21(3)
		&	$<2.5$
		&	0.20(3)
		&	$< 0.12$
		&	11(1)
		&	 $< 0.054$
		&	2.8(2)		\\
\zitem	&	R8520 - Batch 12
		&	Hamamatsu
		&	Bottom array
		&	Gator
		&	9.34
		&	12 pc
		&	mBq/PMT
		&	$<0.70$
		&	$<0.45$
		&	$<15$
		&	$< 0.36$
		&	$<0.68$	&	14(2)
		&	$<0.15$
		&	0.66(7)		\\
\zitem	&	R8520 - Batch 13
		&	Hamamatsu
		&	Top array
		&	LNGS
		&	11.9
		&	10 pc
		&	mBq/PMT
		&	$<0.13$
		&	0.14(3)
		&	$<1.5$
		&	0.38(6)
		&	$< 0.062$
		&	10(1)
		&	$< 0.017$
		&	0.46(5)		\\
\zitem	&	R8520 - Batch 14
		&	Hamamatsu
		&	Bottom array
		&	LNGS
		&	5.12
		&	4 pc
		&	mBq/PMT
		&	$< 0.33 $
		&	$< 0.15 $
		&	$<13$
		&	0.14(7)
		&	$< 0.28$
		&	6(1)
		&	$< 0.13$
		&	0.57(9)		\\
\zitem	&	R8520 - Batch 15
		&	Hamamatsu
		&	Bottom array
		&	Gator
		&	9.51
		&	7 pc
		&	mBq/PMT
		&	$<  0.97$
		&	$<  0.60$
		&	$<22$
		&	$<0.48$
		&	$<0.99$
		&	13(2)
		&	$<0.20$
		&	0.6(2)		\\
\zitem	&	R8520 - Batch 16
		&	Hamamatsu
		&	Bottom array, veto
		&	Gator
		&	5.6
		&	11 pc
		&	mBq/PMT
		&	$<1.1$
		&	0.3(1)
		&	$<21$
		&	$<0.56$
		&	$<0.94$
		&	13(2)
		&	$<0.22$
		&	0.7(1)		\\ 
\zitem	&	R8520 - Batch 17
		&	Hamamatsu
		&	Top/bottom array
		&	LNGS
		&	10.4
		&	5 pc
		&	mBq/PMT
		&	$<0.32$
		&	0.30(7)
		&	$<5.8$
		&	$0.20(5)$
		&	$<0.11$
		&	6(1)
		&	$<0.13$
		&	1.3(1)		\\ 
\zitem	&	QUPIDs
		&	Hamamatsu/ UCLA
		&	R\&D for XENON
		&	Gator
		&	60.0
		&	5 pc
		&	mBq/QUPID
		&	$<0.4$
		&	$0.4(2)$
		&	$< 17$
		&	0.3(1)
		&	$<0.76$
		&	$5.5(6)$
		&	$< 0.21$
		&	$< 0.18 $	\\
\zitem	&	R11410-MOD
		&	Hamamatsu
		&	R\&D for XENON
		&	Gator
	  	&	20.4
		&	1pc
	   	&	mBq/PMT
		&	$<3.8$
		&	$<2.6$
		&	$<95$
		&	$<2.4$
		&	$<4.3$
		&	13(4)
		&	$<1.3$
		&	3.5(6)	\\
\zitem	&	R11410
		&	Hamamatsu
		&	R\&D for XENON
		&	LNGS
	  	&	11.9
		&	1pc
	   	&	mBq/PMT
		&	$<2.7$
		&	3.0(6)
		&	50(20)
		&	6.1(7)
		&	3.1(7)
		&	50(8)
		&	$<0.38$
		&	8.4(8)	\\ \hline
		&	\bf{Connections, \mbox{Cables, etc.}}	&		&		&		&		&		&		&		&		&		&		&		&		\\
\zitem	&	R8520 PMT base
		&	Custom
		&	PMT base
		&	LNGS
		&	6.0
		&	48 pc
		&	mBq/base
		&	0.10(2)
		&	0.07(2)
		&	$<3$
		&	0.16(2)
		&	0.13(3)
		&	$< 0.16$
		&	$< 0.015$
		&	$< 0.010$	\\
\zitem	&	PMMA-PFA \mbox{optical fiber}
		&	Luceat
		&	PMT calibration
		&	ICP-MS
		&
		&	5.0 g
		&	mBq/kg
		&	 6(2)
		&	 6(2)
		&	$<1.9$
		&	$<1.9$
		&	$<0.085$
		&	40(8)
		&	 	\\
\zitem	&	\mbox{Coaxial~cable} (RG174)
		&	Caburn-MDC
		&	PMT Signal
		&	LNGS
		&	5.0
		&	100 m
		&	$\mu$Bq/m
		&	20(10)
		&	$<19$
		&	$<180$
		&	$< 8.9$
		&	$< 14$
		&	200(80)
		&	$<12$
		&	$<3.9$	\\
\zitem	&	\mbox{Kapton~cable} \mbox{(1-CC-0712)}
		&	Caburn-MDC
		&	PMT High Voltage
		&	LNGS
		&	14.0
		&	79.6 g
		&	mBq/kg
		&	$< 8.0 $
		&	$< 11 $
		&	$<350$
		&	$<11$
		&	$<8.4$
		&	610(80)
		&	$< 5.0 $
		&	$< 3.5 $	\\
\zitem	&	Surface mount precision plate, SM5D, 700M$\Omega$ resistors
		&	Japan Finechem
		&	TPC; drift field shaping 
		&	LNGS
		&	20.8
		&	30 pc
		&	mBq/pc
		&	$<0.015$
		&	0.014(3)
		&	0.6(2)
		&	0.027(3)
		&	0.013(4)
		&	0.19(3)
		&	$< 0.004 $
		&	$< 0.003 $	\\ \hline
	&	\bf{Vacuum Parts}	&		&		&		&		&		&		&		&		&		&		&		&		\\
\zitem	&	Ceramic RO4350B prepreg feedthrough
		&	Rogers Corporation
		&	Not used
		&	Gator
		&	4.9
		&	0.014kg
		&	Bq/kg
		&	23(2)
		&	27(3)
		&	$<51$
		&	18(1)
		&	$<2.3$
		&	10(1)
		&	$<0.44$
		&	$<2$	\\
\zitem	&	Ceramic Feedthrough
		&	Caburn-MDC
		&	Electrical feedthrough
		&	LNGS
		&	12.8
		&	0.586 kg
		&	mBq/kg
		&	13(6)
		&	18(6)
		&	210(90)
		&	13(3)
		&	$< 5$
		&	$< 49$
		&	$< 5$
		&	21(2)	\\
\zitem	&	\mbox{Stainless~Steel} Flange
		&	Caburn-MDC
		&	Feedthrough flange
		&	LNGS
		&	22.3
		&	0.495 kg
		&	mBq/kg
		&	7(2)
		&	9(2)
		&	$<83$
		&	7(2)
		&	$<4$
		&	$<36$
		&	$< 4$
		&	6(1)		 	\\ 
\hline
	&	\bf{Environment}	&		&		&		&		&		&		&		&		&		&		&		&		\\
\zitem	&	Concrete
		&	LNGS
		&	LNGS wall
		&	Gator
		&	0.71
		&	0.035 kg
		&	Bq/kg
		&	2.4(8)
		&	3.8(8)
		&	$<160$
		&	15(2)
		&	$<7.2$
		&	42(6)
		&	0.8(2)
		&	$< 0.70$	\\
\zitem	&	Concrete
		&	LNGS
		&	LNGS floor
		&	Gator
		&	0.23
		&	0.033 kg
		&	Bq/kg
		&	7(2)
		&	8(2)
		&	$<190$
		&	26(5)
		&	$<8.5$
		&	170(30)
		&	0.9(3)
		&	 $< 0.58 $	\\ \hline

\end{longtable}
\end{zenumerate}
\setlength{\LTcapwidth}{4in}
\end{landscape}
\normalsize
